\documentclass[11pt]{article}
\usepackage{graphics,latexsym,amsfonts}
\usepackage{graphicx}
\usepackage{float}
\usepackage{picture}
\usepackage[american]{babel}
\usepackage{amsmath,mathtools,amsthm}
\usepackage{enumitem}
\usepackage{color}
\usepackage{comment}
\usepackage{lineno}
\usepackage{algorithm}
\usepackage{algpseudocode}
\usepackage{caption}

\author{
A.\ Pluchino\thanks{Department\ of Physics and Astronomy "Ettore Majorana", University\ of Catania and INFN Sezione di Catania, Italy; alessandro.pluchino@ct.infn.it}$\:$,
A.\ E.\ Biondo\thanks{Dept.\ of Economics and Business, Univ.\ of Catania, Italy; ae.biondo@unict.it}$\:$,
A.\ Rapisarda\thanks{Department\ of Physics and Astronomy "Ettore Majorana", University\ of Catania and INFN Sezione di Catania, Italy; Complexity Science Hub Vienna; andrea.rapisarda@ct.infn.it. \ Paper presented at the Gdansk Summer Solstice conference 2018 on "Discrete models of complex systems" }}

\title{\bf Exploring the role of talent and  luck in getting success}
\begin{document}
\date{ }
\maketitle

\begin{abstract}
We review recent numerical results on the role of talent and luck in getting success by means of a schematic  agent-based model \cite{tvl}. In general the role of luck is found to be very relevant in order to get success, while talent is  necessary but not sufficient.  Funding strategies to improve the success of the most talented people are also discussed.

\medskip
\medskip
\noindent \textbf{Keywords:} Success, Talent, Luck, Randomness, Serendipity, Funding strategies.
%\medskip
%\noindent \textbf{\textsf{JEL} Classification:}.
\end{abstract}

\vskip 2cm

Power-law distributions are ubiquitous in many physical, biological and  socio-economical complex systems and are a sort of mathematical signature of  strong correlations and  scale invariant hierarchical  structure \cite{Barabasi,Newman,Tsallis}. It was the economist Pareto  the first one to show the presence of  power-law distributions in the wealth of countries and of single individuals \cite{Pareto}. This fact  indicates  a strong inequality in our society: a very small amount of people have the same richness of the rest of the world.  In some sense one could consider the personal wealth as a proxy of success, and think that a very successful person should be also, proportionally, a very talented individual. But this point of view, characteristic of the standard meritocratic paradigm, is in strict contrast with the accepted evidence that human features and qualities (heigth, IQ, weight, etc.), and also efforts (evaluated, for example, in working hours), follow a symmetric Gaussian distribution around a given mean: actually, there is not an individual who is thousands of times more talented or more skilled or more intelligent than the rest of the population, just as there is not an individual who works thousands of times more than another one. 

A key to understand this apparent contradiction can be found in the structure and in the complexity of our globally networked socio-economic system, full of feedback mechanisms and winners-take-all domains. In this highly non-linear context, the adoption of a simple linear paradigm to connect intellectual capacity or productivity efforts with the scale invariant wealth distribution does result at least rather naive. Indeed, it frequently happens that small advantage/disadvantage in IQ or small differences in efforts could lead to large increase/decrease in the resulting income, since the latter may be strongly influenced by cumulative effects induced by the multiplicative dynamics of the system. In such a context, so sensitive to external circumstances, it can also happen that some small random and unpredictable event, completely independent of talent and efforts, may provide the seed for generating a cascade process of lucky opportunities which end to generate a final power-law distribution of success or wealth. 

The fundamental role of luck/chance in our life, as well as that of unpredictable events not under our control, has been, traditionally, strongly underestimated. This fact has been recently realized and discussed by authors like Taleb \cite{Taleb1,Taleb2}, Mauboussin \cite{Mauboussin}, Frank \cite{Frank} and Watts \cite{Watts}. On the other hand, there is a lot of literature presenting data in favour of the importance of chance in getting success. A few examples among many others are the following: a) scientists have the same probability along their career of publishing their most important paper  \cite{Sinatra}; b) individuals with earlier surname initials are significantly more likely to receive tenure positions \cite{Einav}; c) one's position in an alphabetically sorted list may be important in determining access to over-subscribed public services \cite{Jurajda}; d)  people with easy-to-pronounce names are judged more positively \cite{Laham}; and even the probability of developing a cancer is often due to random errors in DNA replication \cite{Tomasetti}.

\begin{center} 
* * *
\end{center}

In a recent paper \cite{tvl}, by means of an agent-based model, we tried to quantify in a simple but realistic way the respective role of luck and talent in order to have a successful career. We summarize the main results in the following. 

The model simulate the evolution of careers of a group of N agents (N=1000) over a working period of 40 years. Agents are endowed with a talent $T_i \in[0,1]$, extracted from a Gaussian distribution \cite{Stewart} centered at 0.6 and with a standard deviation 0.1, and have the same initial capital/success $C_i = 10$. They are placed at random in fixed positions within a virtual squared world and are surrounded by a certain number $N_E$  events, someone lucky, someone else unlucky, moving at random during each simulation run.
 
The initial capital of the agents can change every six months according to these simple rules: 

(1) If a lucky event intercepts the position of agent $A_k$, this means that a lucky event has occurred during the last six month; as a consequence, agent $A_k$ doubles her capital/success with a probability proportional to her talent $T_k$. In other words, $C_k(t)=2C_k(t-1)$ only if $rand[0,1] < T_k$, i.e. if the agent is smart enough to profit from his/her luck.       

(2) If an unlucky event intercepts the position of agent $A_k$, this means that an unlucky event has occurred during the last six month; as a consequence, agent $A_k$ halves her capital/success, i.e. $C_k(t)=C_k(t-1)/2$.

We discuss  in the following the main results of the model, presenting numerical simulations averaged over $100$ runs (events) with different initial conditions. 

In panel (a) of Figure 1, the tail of the global distribution of the final capital/success for all the agents collected over the $100$ events is shown in log-log scale. The numerical data are well fitted by a  power-law  with a slope equal to $-1.33$: a scale invariant behavior of capital and the consequent strong inequality among individuals, consistent with the Pareto's "80-20" rule, is therefore observed.   
In panel (b), we show the final capital of the most successful individuals only, for each one of the $100$ events, reported as function of their talent. The highest capital $C_{best}=40960$ was obtained by an agent with a talent $T^*=0.6048$, practically equal to  the mean of the talent distribution ($m_T=0.6$). On the other hand, the most talented among the most successful individuals (with a talent $T_{max}=0.91$) accumulated at the end of her career a capital  $C=2560$, equal to only $6\%$ of the highest one.

%%%%%%%%%%%%%%%%  FIG. 1 %%%%%%%%%%%%%%%%%%
\begin{figure}[t]
\begin{center}
\includegraphics[width=5.in,angle=0]{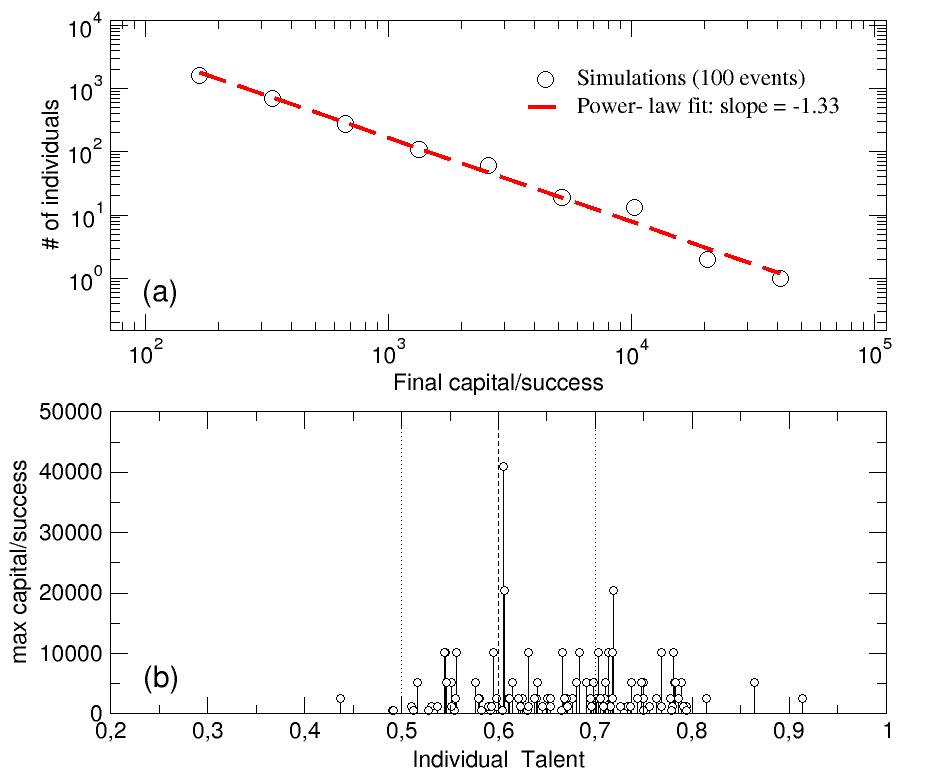}
\caption{\small 
Panel (a): Tail of the distribution of the final capital/success of the agents calculated over $100$ events  and considering  different random initial conditions. We show also a Pareto-like power-law fit  with a slope equal to  $-1.33$. 
Panel (b): Here we show the final capital of the most successful individuals in each of the $100$ events  as function of their talent. People with a medium-high talent result to be, on average, more successful than people with low or medium-low talent, but very often the most successful individual is a moderately gifted agent and only rarely the most talented one. The mean value of the talent distribution $m_T$, together with the values $m_T \pm \sigma_T$, are also reported as vertical dashed and dot lines respectively.}
\label{talent-success} 
\end{center}
\end{figure}
%%%%%%%%%%%%%%%%%%%%%%%%%%%%%%%%%%%%%%%%%%%%

From these simulations and others shown in the original paper \cite{tvl}, our model seems able to account for many of the features characterizing the largely unequal distribution of richness and success observed in our society.
The results of the model also show, in quantitative way, that having a great talent is not a sufficient condition to guarantee a successful career. On the other hand, people with a talent slightly above the average, provided they have been very lucky, are often able to reach the top of success, a fact which is frequently observed in real life \cite{Taleb1,Taleb2,Frank}. Thus, it seems that luck/chance does play an important role in reaching a very successful position and this evidence poses a fundamental question about meritocracy in our society.

\begin{center} 
* * *
\end{center}

The meritocratic criteria used to assign honors, funds or rewards are often based on personal wealth or success of individuals, being their talent, in many general contexts, not easy to be evaluated. Our findings strongly suggest that those particular individuals could have been, at the end of the story, just the most lucky. What's worse, such strategies can eventually exert a further reinforcing action on the luckiest individuals through a kind of positive feedback mechanism, the famous "rich get richer" process (also known as  "Matthew effect" \cite{Merton1,Merton2}), with a more unfair result.

Just to give an example, in the field of research funding, recent studies  \cite{Fortin,Mongeon,Jacob} found that the most funded research groups do not stand out in terms of output and scientific impact, suggesting that it is more productive to follow funding strategies that foster "diversity" rather than "excellence". On the other hand, if chance matters as we support, it should not be strange that meritocratic strategies are less effective than expected, in particular when one evaluates merit \textit{ex-post}. After all, the word "serendipity" is commonly used for those unexpected discoveries made by chance \cite{Merton,Murayama}. Going from penicillin to graphene \cite{tvl}, there is a long anecdotical list of discoveries just due to lucky opportunities. 
That is why it is quite important to support  curiosity-driven research, being  very difficult to predict the final outcomes of a research project. We already addressed the problem of "naive meritocracy" in several papers, showing the effectiveness of strategies based on random choices in management, politics and finance \cite{Pluchino1,Pluchino2,Pluchino3,Biondo1,Biondo2,Biondo3,Biondo4,Biondo5}. In the following we discuss how it is possible, in the context of the model presented here, to increase the minimum level of success of the most talented people in a world where luck/chance plays an important role. 

Let us imagine to periodically distribute a funding capital $F_T$  among the agents following different criteria. In ref. \cite{tvl} we compared several distribution strategies in order to provide additional resources that could allow the most talented agents to increase their initial capital. We  assumed to distribute a fixed capital $F_T=80000$ every $5$ years, during a period of $40$ years spanned by each simulation run, so that $F_T/8$ units of capital will be allocated from time to time. We used as an indicator to check the effectiveness of the adopted funding strategy: the number $N_T$ (averaged over the $100$ simulation events) of individuals with talent $T$ greater than 1 standard deviation and with a final success/capital greater than the initial one (we checked that is is a robust measure).

%%%%%%%%%%%%%%%%  FIG. 2 %%%%%%%%%%%%%%%%%%
\begin{figure}[t]
\begin{center}
\includegraphics[width=5.in,angle=0]{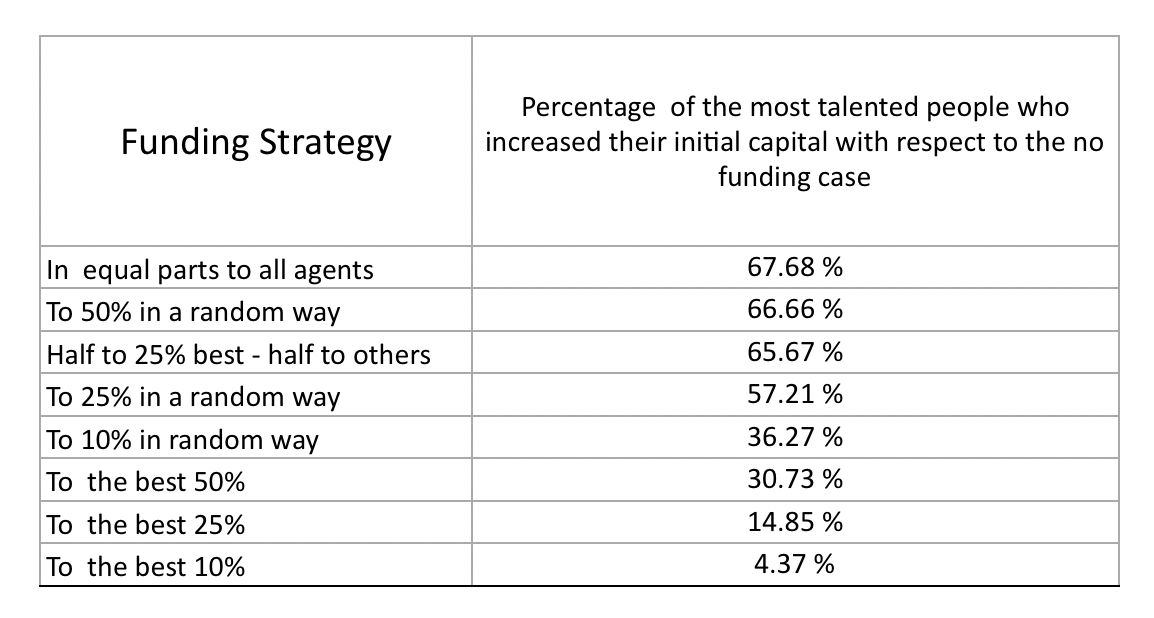}
\caption{\small 
Comparison among different funding strategies averaged over 100 events. A total funding capital of 80000 units was distributed among the agents every 5 years in a period of 40 years. We report for each strategy the final percentage of the most talented agents (those with T greater than one standard deviation with respect to the mean) who were able to increase their initial capital, compared with the no funding case. See text for further details.}
\label{table} 
\end{center}
\end{figure}
%%%%%%%%%%%%%%%%%%%%%%%%%%%%%%%%%%%%%%%%%%%%

Considering the percentage of these agents with respect to the case with no funding, we can compare the results of each adopted strategy in order see which one is the most effective. Some results are reported in Figure 2. For  more detail please refer to the original paper \cite{tvl}. 
Looking at the table shown in the figure, if the goal is to have the most talented persons with a final capital greater than the initial one, it is much more convenient to distribute periodically equal amounts of capital to all individuals rather than giving  a greater capital only to a small percentage of them, selected through their actual level of success - already reached at the moment of the distribution. 
On one hand, the table shows that the most "egalitarian" strategies, which assigns an equal amount  of capital every 5 years to all the individuals, is the most efficient way to distributed funds.
On the other hand, the most "elitarian" strategies which assign every 5 years funds only to the best $50\%$, $25\%$ or even $10\%$ of the already successful individuals, are all at the bottom of the ranking in all of these cases.
If one also considers psychological factors (not taken into account in the simulations but relevant in the real world), a mixed strategy could be a good compromise with respect to the egalitarian one.          
Finally, looking again at the funding strategy table, it is also worthwhile to stress the counterintuitive  high efficiency of the random strategies, which occupy two out of  the three best scores in the general ranking.     

In ref.\cite{tvl} we studied also the importance of the environment or of the education in order to improve the success of the most talented agents. We saw that  a stimulating environment, richer  of opportunities, associated to an appropriate strategy for the distribution of funds and resources, are important factors in exploiting the potential of the most talented people, giving them more chances of success with respect to the moderately gifted, but luckier, ones. At the macro level, any policy able to influence those factors and to sustain talented individuals, will have the result of ensuring collective progress and innovation.         

\begin{center} 
* * *
\end{center}

In summary, we have shown, by means of an agent-based model, how it is possible to quantify the role of talent and luck in order to reach success,  starting from very simple assumptions. Our simulations show that, although talent is normally distributed among agents, the final distribution of success/capital follows a power-law behavior similar to the Pareto law observed in the real world. We have also found that the most successful agents are almost never the most talented ones, but just very lucky individuals with a medium level of talent, another feature often perceived in real life. The model thus shows the importance, very frequently underestimated, of lucky events in determining the final degree of individual success.  We have also compared different  funding strategies to increase the level of success of the most talented agents, finding that the  most egalitarian ones are those which are the most effective in this respect.


\begin{thebibliography}{99}
	
\bibitem{tvl} A. Pluchino, A.E. Biondo, A. Rapisarda, "Talent vs. luck: the role of randomness in success and failure". Advances  in Complex Systems, Vol. 21, Nos. 3 \& 4 (2018) 1850014.

	

\bibitem{Barabasi}
A. L. Barab\'asi, R. Albert. "Emergence of Scaling in Random Networks". Science, Vol. 286, Issue 5439, pp. 509-512 (1999). 
	
\bibitem{Newman}
M. E. J. Newman. "Power laws, Pareto distributions and Zipf's law". Contemporary Physics, 46 (5): 323-351 (2005).	

\bibitem{Tsallis}
C. Tsallis. "Introduction to Nonextensive Statistical Mechanics. Approaching a Complex World". Springer (2009).	
	
\bibitem{Pareto}
V. Pareto. Cours d'Economique Politique, vol. 2 (1897).


\bibitem{Taleb1}
N. N. Taleb. "Fooled by Randomness: The Hidden Role of Chance in Life and in the Markets". London, TEXERE (2001).

\bibitem{Taleb2}
N. N. Taleb. "The Black Swan: The Impact of the Highly Improbable". Random House (2007).

\bibitem{Mauboussin}
M. J. Mauboussin. "The Success Equation: Untangling Skill and Luck in Business, Sports, and Investing". Harvard Business Review Press (2012).

\bibitem{Frank}
R. H. Frank. "Success and Luck: Good Fortune and the Myth of Meritocracy". Princeton University Press, Princeton, New Jersey (2016). 

\bibitem{Watts}
D. J. Watts. "Everything Is Obvious: Once You Know the Answer". Crown Business (2011).



\bibitem{Sinatra}
R. Sinatra, D. Wang, P. Deville, C. Song and A.-L. Barabási. "Quantifying the evolution of individual scientific impact". Science 354, 6312 (2016).

\bibitem{Einav}
L. Einav and L. Yariv. "What's in a Surname? The Effects of Surname Initials on Academic Success". Journal of Economic Perspective, Vol. 20, n. 1, p.175 188 (2006). 

\bibitem{Jurajda}
S. Jurajda, D. Munich. "Admission to Selective Schools, Alphabetically". Economics of Education Review, Vol. 29, n. 6, p.1100-1109 (2010).

\bibitem{Tilburg}
W. A. P. Van Tilburg, E. R. Igou. "The impact of middle names: Middle name initials enhance evaluations of intellectual performance". European Journal of Social Psychology, Vol. 44, Issue 4, p.400-411 (2014).

\bibitem{Laham}
S. M. Laham, P. Koval, A. L. Alter. "The name-pronunciation effect: Why people like Mr. Smith more than Mr. Colquhoun". Journal of Experimental Social Psychology 48, p.752-756 (2012).


\bibitem{Tomasetti}
C. Tomasetti, L. Li, B. Vogelstein. "Stem cell divisions, somatic mutations, cancer etiology, and cancer prevention". Science 355, 1330-1334 (2017).
 

\bibitem{Stewart}
J. Stewart, "The Distribution of Talent". Marilyn Zurmuehlin Working Papers in Art Education 2: 21-22 (1983).


\bibitem{Merton1}
R. K. Merton. "The Matthew effect in science". Science 159, 56-63 (1968).

\bibitem{Merton2}
R. K. Merton. "The Matthew effect in science, II: Cumulative advantage and the symbolism of intellectual property". Isis: A Journal of the History of Science 79, 606-623 (1988).

\bibitem{Fortin}
J.-M. Fortin, D. J. Curr. "Big Science vs. Little Science: How Scientific Impact Scales with Funding". PLoS ONE 8(6): e65263 (2013).

\bibitem{Mongeon}
P. Mongeon, C. Brodeur, C. Beaudry et al. "Concentration of research funding leads to decreasing marginal returns". Research Evaluation 25, 396-404 (2016).

\bibitem{Jacob}
B. A. Jacob, L. Lefgren."The impact of research grant funding on scientific productivity". Journal of Public Economics 95 (2011) 1168-1177.

\bibitem{Pluchino1} 
A. Pluchino, A. Rapisarda, and C. Garofalo. "The Peter principle revisited: A computational study". Physica A: Statistical Mechanics and its Applications, 389(3):467-472 (2010).

\bibitem{Pluchino2}
A. Pluchino, C. Garofalo, A. Rapisarda, S. Spagano, and M. Caserta. "Accidental politicians: How randomly selected legislators can improve parliament efficiency". Physica A: Statistical Mechanics and Its Applications, 390(21):3944-3954 (2011).

\bibitem{Pluchino3}
A. Pluchino, A. Rapisarda, and C. Garofalo. "Efficient promotion strategies in hierarchical organizations". Physica A: Statistical Mechanics and its Applications, 390(20):3496-3511 (2011).

\bibitem{Biondo1}
A. E. Biondo, A. Pluchino, A. Rapisarda, D. Helbing. "Reducing financial avalanches by random investments". Phys. Rev. E 88(6):062814 (2013).

\bibitem{Biondo2}
A. E. Biondo, A. Pluchino, A. Rapisarda, D. Helbing. "Are random trading strategies more successful than technical ones". PLoS One 8(7):e68344 (2013)

\bibitem{Biondo3}
A. E. Biondo, A. Pluchino, A. Rapisarda. "The beneficial role of random strategies in social and financial systems". J. Stat. Phys. 151(3-4):607-622 (2013).

\bibitem{Biondo4}
A. E. Biondo, A. Pluchino, A. Rapisarda. "Micro and macro benefits of random investments in financial markets". Cont. Phys. 55(4):318-334 (2014).

\bibitem{Biondo5}
A. E. Biondo, A. Pluchino, A. Rapisarda. "Modeling financial markets by self-organized criticality". Phys. Rev. E 92(4):042814 (2015).


\bibitem{Merton}
R. K. Merton, E. Barber. "The Travels and Adventures of Serendipity". PUPPrinceton (2004).

\bibitem{Murayama}
K. Murayama et al.. "Management of science, serendipity, and research performance". Research Policy 44 (4), 862-873 (2015).


\end{thebibliography}
\end{document}